\documentclass{acm_proc_article-sp}
\usepackage{graphicx}

\begin{document}

\conferenceinfo{}{Bloomberg Data for Good Exchange 2016, NY, USA}

\title{Making Public Safety Data Accessible \\in the Westside Atlanta Data Dashboard}

\numberofauthors{3}
\author{
\alignauthor
Katie O'Connell\\
Yeji Lee\\
Firaz Peer\\
       \affaddr{Georgia Institute of Technology}\\
       \affaddr{Atlanta, GA}\\
       \email{l.katieoconnell@gmail.com}\\
       \email{yejilee7@gatech.edu}\\
       \email{firazpeer@gatech.edu}\\
\alignauthor
Shawn M. Staudaher\\
       \affaddr{University of Wyoming}\\
       \affaddr{Laramie, WY}\\
       \email{sstaudah@uwyo.edu}
\and
\alignauthor Alex Godwin\\
Mackenzie Madden\\
Ellen Zegura\\
       \affaddr{Georgia Institute of Technology}\\
       \affaddr{Atlanta, GA}\\
       \email{alex.godwin@gatech.edu}\\
       \email{mmadden6@gatech.edu}\\
       \email{ewz@cc.gatech.edu}\\
}

\maketitle
\begin{abstract}
Individual neighborhoods within large cities can benefit from independent analysis of public data in the context of ongoing efforts to improve the community. Yet existing tools for public data analysis and visualization are often mismatched to community needs, for reasons including geographic granularity that does not correspond to community boundaries, siloed data sets, inaccurate assumptions about data literacy, and limited user input in design and implementation phases.  In Atlanta this need is being addressed through a Data Dashboard developed under the auspices of the Westside Communities Alliance (WCA), a partnership between Georgia Tech and community stakeholders. In this paper we present an interactive analytic and visualization tool for public safety data within the WCA Data Dashboard. We describe a human-centered approach to understand the needs of users and to build accessible mapping tools for visualization and analysis. The tools include a variety of overlays that allow users to spatially correlate features of the built environment, such as vacant properties with criminal activity as well as crime prevention efforts. We are in the final stages of developing the first version of the tool, with plans for a public release in fall of 2016.
\end{abstract}





\section{Introduction}
Individual neighborhoods within large cities can benefit from independent analysis of public data in the context of ongoing efforts to improve the community. For example, communities in the Westside of Atlanta have been changing their neighborhoods for the better by organizing amongst themselves and collaborating with organizations that have a local presence. Georgia Tech has partnered with Westside communities via the Westside Community Alliance (WCA), a communications network started in 2011 by the Ivan Allen College of Liberal Arts in collaboration with the College of Design and the Office of Government and Community Relations. The WCA works to build and sustain relationships among constituencies located in West Atlanta to strengthen partnerships around issues of common concern. 

Since its inception in 2011, the WCA recognized the importance of data dissemination as a tool for community development. In February 2016 the WCA launched the WCA Data Dashboard.  This on-line website is designed to be a one-stop data shop with information presented in locally recognized and meaningful geographies.  Rather than census tracts or zip codes, data is presented by neighborhood or Neighborhood Planning Unit (NPU). In the Atlanta area, neighborhoods and NPUs represent the unit of local community organization as well as local identity and pride. The Dashboard developers have gathered previously siloed data sets and integrated them into one platform to support examining data relationships. Central to the Dashboard's design is regular engagement with local organizations and community groups both for design feedback and for data literacy training. The Data Dashboard is organized into portals that correspond to community concerns. Currently there are five portals---community profile, education, historic data, historic timeline, and resource library---with several more in the pipeline~\footnote{http://wcadatadashboard.iac.gatech.edu}.

In Summer 2016, the WCA served as a client and partner in the Atlanta Data Science for Social Good (DSSG-ATL) summer internship program. DSSG-ATL is modeled after the DSSG summer fellowship program started at the University of Chicago in 2013; students work full-time in teams on projects that come from local partners, with a focus on supporting understanding and decision-making based on data, in domains of social importance. DSSG-ATL started in 2014~\footnote{http://www.dssg-atl.io}. 

In this paper, we describe our participatory approach to building the public safety module of the WCA's Data Dashboard, with some key insights into how one can approach similar projects in the future. We conducted interviews with key community stakeholders and participated in local government meetings to understand the needs of our users. These needs include the ability to locate where crimes have occurred, a metric to quantitatively evaluate the efficiency of their public safety programs, and a means to learn how other factors such as education, age, transportation, housing, and more are related to crime. To satisfy these needs, we introduce a mapping tool with the capability to locate current and historic hotspots of criminal activity. This tool includes a variety of overlays that allow users to spatially correlate features of the built environment, such as code violations  with criminal activity as well as crime prevention efforts. We also present a statistical model that highlights correlations between crime and other socio-economic factors specific to particular neighborhoods in Atlanta.

Parts of the public safety module have been built with Tableau and web-based mapping libraries like CartoDB and Leaflet.js to help maintain consistency with the existing Data Dashboard. We are in the final stages of developing the first version of the tool, with plans for a public release in the fall of 2016.

This paper is organized as follows. In Section~\ref{sec:related} we review related work on public mapping systems, participatory design, and public data use, along with design goals of the WCA Data Dashboard to highlight how these goals are not fully met by existing work. We then describe our approach to gathering user input and understanding user needs. The public safety portal design is described in Section 4 and includes a description of the data sets used to populate the portal. Section 5 contains a discussion of issues that arose during portal development and that suggest the need for additional work in the public data community. We conclude in Section 6.

\section{Related Work and Design Goals}
\label{sec:related}
We divide related work into three categories: public geographical information systems; participatory design; and community-based mapping tools specific to crime. We then list the design goals of the Data Dashboard in general and the Public Safety portal in particular.

\subsection{Public Geographical Information}
Community participation in urban planning is an area of growing interest for urban developers and researchers in geographical information science (GIS)~\cite{dunn2007participatory}. Visualization tools in GIS encompass a wide scale, ranging from traditional pen-based sketching approaches to advanced computerized mapping toolkits~\cite{al2002visualization}. While it is easy to imagine that advancements in GIS alone can serve to empower community members due to more powerful and easier-to-use spatial analysis systems, the role that GIS software plays in community development can vary dramatically. For example, when GIS is used only to inform members of the community of new policy without inviting participation, it can serve to further disenfranchise grassroots organizations and local stakeholders. This one-sided definition of participation can be classified as educational at its most optimistic or manipulative at its most cynical~\cite{schlossberg2005delineating}.

In contrast, public participation geographic information systems (PPGIS) attempt to make GIS tools more widely available and facilitate interaction with all stakeholders~\cite{sieber2006public} . Approaches such as Bottom-up GIS~\cite{talen2000bottom} allow members of the community to participate in the planning of urban growth by incorporating their perception of the neighborhood into analysis of spatial data (e.g., areas where a large number of crimes are perceived to be reported but never recorded in the official public data). In our research, we have attempted to incorporate community member perception into the design of a tool that correlates traditional measures of public safety (e.g., crime reports) with what local residents see as the most important assets affecting public safety through community development and outreach programs (e.g., churches, schools). Tools using this approach can themselves become inroads to improving the participation of young people in neighborhood planning by capturing the more qualitative aspects of youth perception of a neighborhood  and facilitating the design of subsequent outreach programs~\cite{dennis2006prospects}. 

\subsection{Participatory Design}
The idea of talking to users to elicit their needs is not a new one, and has been formalized into a general set of principles under Human/User Centered Design, User Driven Development, and Participatory Design~\cite{simonsen2012routledge}. While the specific methods employed depend on the domain and the kinds of users, the general idea is to talk with community members early in the design process to determine their needs via focus groups, interviews, or site visits. User-centered design is well-suited for PPGIS, and provides a framework for establishing the needs of community members in learning about and utilizing GIS within the context of neighborhood planning~\cite{haklay2003usability}. The WCA has been working with the Westside community for years, and is well versed with the needs of the community. Our weekly meetings with them helped us better understand the people and data we were working with and also prioritize the features we hoped to include. 

\subsection{Crime Mapping Tools}
Community-based mapping tools often include straightforward depictions of crime locations on a map so that police departments can provide greater transparency of public safety (e.g., the Socrata platform~\cite{socrata}). Online news sources, such as the Chicago Tribune \footnote{http://crime.chicagotribune.com/}, have published crime data and maps broken up by neighborhood and crime type. The Atlanta Journal Constitution~\footnote{http://crime.myajc.com/} has used publicly available crime data from the Atlanta Police Department to produce a site that maps trends in crime since 2009 by police zones and beats. This type of transparency and open data can yield to more in-depth analysis, such as the Million Dollar Blocks project, in which the location and cost of the NYC incarcerated is correlated with the location that they originate from rather than the location in which the crime occurred~\cite{flink2009million}. In a more widespread fashion, the MIT Media Lab's Data USA tool provides general demographic data and public safety information across the country by combining multiple open sources into a single location-based form~\cite{lohr}. While these and other tools offer overviews of crime and public safety in a city or neighborhood, we could not find existing participatory systems that would help residents visualize the impact of public safety efforts or programs in their neighborhoods, which was a concern that frequently came up in our discussions with community members and the WCA.

\subsection{Design Goals}
Because the Data Dashboard was developed within the WCA, it shares design goals and values with the overarching organization. In particular, the WCA is a partnership between a university (Georgia Tech) and local communities. The structure and priorities of the WCA are arrived at via community engagement. Staff in the WCA live in these communities and are regular participants in community meetings and events. Long-standing relationships with the community are created and sustained through regular and myriad interactions. The Data Dashboard was designed to be accessible to citizens in ways that fit how citizens think about their communities, useful for citizen information gathering and advocacy, and integrated so that citizens do not need to navigate and synthesize data from disparate sources. 

For example, would it be possible for churches to input the different youth programs they run, into a tool to determine how it has impacted juvenile crime in their neighborhoods? As part of Operation Shield, the Police Foundation, with the help of a \$1.2 million grant from Invest Atlanta, has recently installed about 80 cameras in different parts of the neighborhood that are considered hotspots for drug related crimes. Can we overlay these camera locations on a map to see what impact they may have had on these hotspots? The Westside also has a high number of vacant properties, which many residents believe is the cause for high crime in their neighborhoods. Can we numerically determine the relationship between crime and code violations, so residents can make a case with the city to demolish specific properties that have a high correlation with crime? These questions are representative of the kinds we hope we can answer through our tool. But before we could build any of it, our first task was to understand the people we were working with, to make sure that the tool we build is the one they need. Community feedback and design participation---described in more detail in the next section---are key to achieving accessibility and usefulness.

\section{Approach}
\label{sec:approach}
In addition to weekly meetings between the WCA the DSSG team, we also participated in NPU meetings. These are public planning meetings attended by residents and other interested  stakeholders within the community. The meetings provide a place for residents to interact with community and city leadership. These tend to be highly contested spaces, as committee members share updates, residents hear about and vote on specific changes they would like to see implemented, and organizations/researchers get buy-in for various initiatives they have in the pipeline. Attending these NPU meetings gave us a good sense of the issues residents were currently grappling with as well as a first-hand exposure to the dynamics of community governance. 

Some high priority issues have their own committees and meeting schedules; public safety has recently emerged as an area worthy of committee instantiation. The Vine City Public Safety Committee meets on a regular basis to discuss the status of the many public safety programs, and their meetings are attended by police officers and residents alike. We attended one instance of this meeting, where we described our preliminary tool, along with some screenshots of the kinds of visualizations we had in mind, to seek feedback.

The attendees at the public safety meeting seemed excited about the prospect of having free and open access, along with the ability to analyze crime and code violations data within their neighborhoods. We spoke with one of the code enforcement officers present at the meeting, who was interested in talking with us further about ways his department could use our data visualizations in their day-to-day operations. This was significant, as public safety officers were one of the user groups we were hoping to design for. He introduced us to one of the senior analysts at the Code Enforcement division of the Atlanta Police Department, who gave us an overview of exactly how his team goes about collecting and reacting to code violation complaints in their jurisdiction. This meeting also helped clarify many questions we had about the code violations dataset we were working with.


\section{Public Safety Portal}
\subsection{Data Sets}
\label{sec:data}
The crime data contains all reported crime in the City of Atlanta from 2008--2015 and was provided to the WCA via a Freedom of Information Act Request. This data set contains 875,491 records, with relevant columns including report date, occurrence date, postal address, the Uniform Crime Reporting Code (the type of crime committed), latitude, and longitude. Of these records, 9,712 did not contain latitude and longitude, and were geocoded using the arcGIS geocoding API with the provided postal address. Also, these data were spatially joined with the boundaries of NPUs and the neighborhoods of Atlanta.

The code violation data was provided to the WCA through a request to the Atlanta Police Department Office of Code Enforcement. This data set contains all code violation cases in the City of Atlanta from 2011 to early 2016, totaling 42,102 cases. Relevant columns include report date, latest inspection date, postal address, the status of the case, and details about the property including if the house is open and vacant, the presence of overgrowth, active utilities, and more. These data did not contain latitude and longitude, but these were added with the arcGIS geocoding API.

The 2010 Census data was taken from the Neighborhood Nexus, a data portal for the City of Atlanta. These data are broken down by neighborhood and NPU, with a wide variety of parameters including demographics, education, housing, economic information, and more.

It was important both to the WCA and to neighborhood advocacy groups that the public safety portal not just highlight the negative aspects of the community so the page includes a number of assets like schools, religious institutions, parks, and transit stops.  This data was downloaded from the Atlanta Regional Commission's Open Data portal and included latitude and longitude information. 

\subsection{Portal Design}
Based on our discussions with WCA and our observations from NPU meetings, the public safety dashboard contains three sections: an aggregate page, a correlations page, and a mapping tool.

The aggregate module and the correlations module are designed to support two broad use cases. The first is to either confirm or dispel previously held beliefs about public safety. The second is to provide a tool to help the residents develop their own community-led public safety programs by understanding the contexts and trends of crime and code violations.

\begin{figure}[htb]
\centering
\includegraphics[width=\columnwidth]{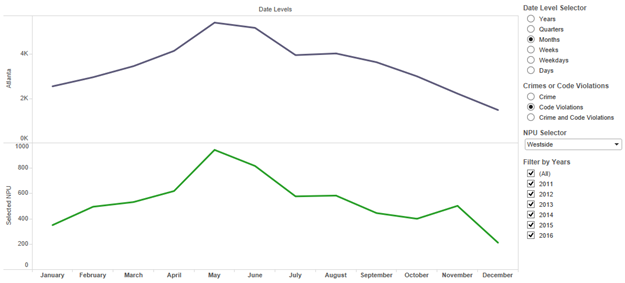}
\caption{Aggregated Module - Crime data aggregated by month for the city of Atlanta (top) and the Westside (bottom).}
\label{crime-by-month}
\end{figure}

The aggregate module contains general overview information about public safety in Atlanta and allows the users to compare Atlanta against their specific NPU of interest. The first component on the aggregate page is a time view (see Figure~\ref{crime-by-month}). The user is able to select a time frame (years, quarters, months, weeks, weekdays, or days) and compare the trends in crime and code violations for Atlanta and a selected NPU. One of the options for the NPU Selector is ''Westside'' which includes NPUs K, L, and T. They can choose to see only crime, only code violations, or both. This time view was frequently requested by the community and from city officials including the Atlanta Police Department Office of Code Enforcement. Common questions include how specific timespans (e.g., football season) or days of the week affect the numbers of crimes or code violations. Organized neighborhood watch groups can utilize this information to plan and strategize accordingly.

\begin{figure}[htb]
\centering
\includegraphics[width=\columnwidth]{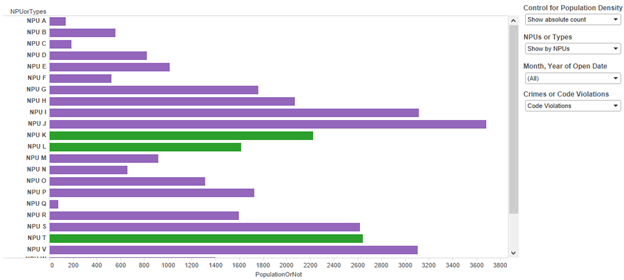}
\caption{Aggregated Module - The number of code violations for NPUs A to V in the City of Atlanta. The Westside NPUs are highlighted in green.}
\label{all-npus}
\end{figure}

The second component is an interactive bar graph with several options for general public safety profile (Figure~\ref{all-npus}). The user is able to select between crimes and code violations and may control for population. The three NPUs that comprise the Westside are highlighted in green for an easy comparison to the other NPUs of Atlanta in purple. The month-year filter allows the users choose a time frame. This component is a broader report meant for city officials such as the Atlanta Police Department require an easy to understand monthly breakdown of crimes and code violations per type and per NPU.

\begin{figure}[htb]
\centering
\includegraphics[width=\columnwidth]{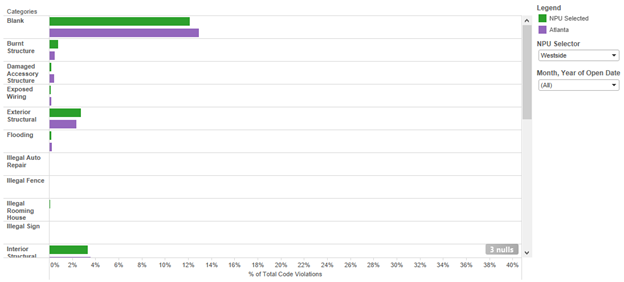}
\caption{Aggregated Module - Percent of total code violations for the Westside (green, upper) and the City of Atlanta (purple, lower).}
\label{npu-and-atlanta}
\end{figure}

The final component of the aggregate module highlights differences between a particular neighborhood and the rest of the City of Atlanta (see Figure~\ref{npu-and-atlanta}). This is done by dividing crime, or code violations, into types and then calculating the percentage of the total for each type. A user such as a resident, business owner, or community partner may learn that which types of crime are less prevalent in their neighborhood in comparison to the rest of Atlanta.

\begin{figure}[htb]
\centering
\includegraphics[width=\columnwidth]{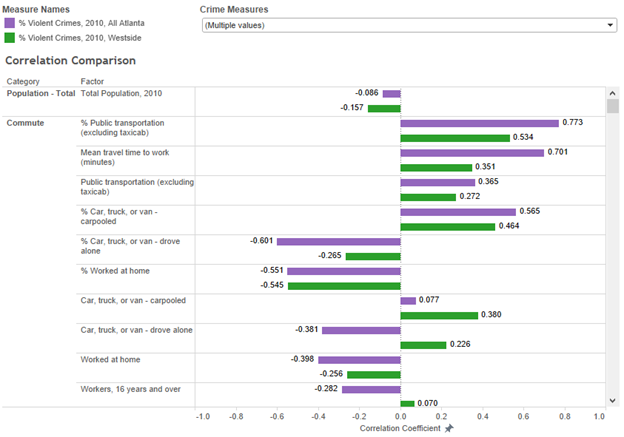}
\caption{Correlations Module - Correlation coefficients between Population, and Commute factors with violent crime in the City of Atlanta and the Westside.}
\label{correlation-coefficients}
\end{figure}

The second correlations module highlights relationships between various factors from from the 2010 census (see Section~\ref{sec:data}) and various crimes taken across each neighborhood in the City of Atlanta. In Figure~\ref{correlation-coefficients} the crime measures selected are the percentage of violent crime (robbery, murder, rape, and aggravated assault) with a comparison between the Westside and the City of Atlanta. This page allows users to answer questions such as, ``Is the percentage of senior citizens in a neighborhood correlated with crime? What specific types of crime are correlated?'' or ``Are vacant houses correlated with crime?'' Although correlation does not necessarily imply causation, this module provides a starting point to answer long-held questions.

\begin{figure*}[ht]
  \includegraphics[width=\textwidth]{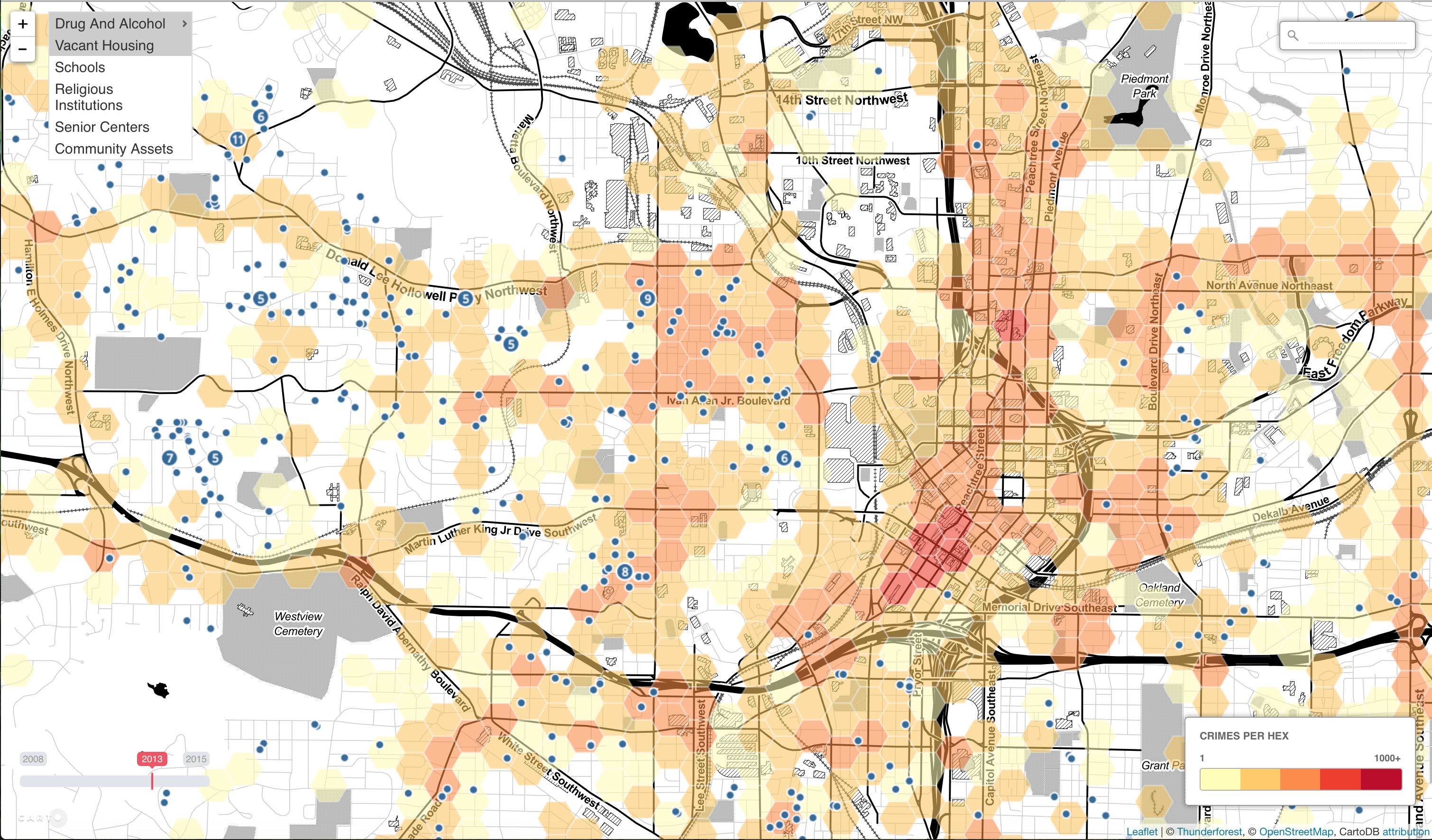}
  \caption{Map Module - The map displays drug and alcohol related crime (colored hexes) and vacant houses selected (blue circular pins).}
  \label{fig:mapping-tool}
\end{figure*}

The third module in the dashboard is the spatial visualization of the crime data, code violations, and community assets. The primary goal of these visualizations is to provide users with a means to precisely pin-point where crimes occur, and to give them the ability to learn how these locations change by crime-type and over time. Additionally, the map allows users to focus on specific geographic areas to examine the relationship between crime, code violations, and community assets. 

The crime data is visualized with a hex-based heat map (Figure \ref{fig:mapping-tool}). This map is constructed by counting the number of crimes within each hex and assigning one of five colors based on a logarithmic scale (i.e. the first color represents a single crime per hex, the second color represents 2-10 crimes per hex, the third color is 11-100 crimes per hex, and so on). This logarithmic aggregation of crime into colored hexes naturally highlights hot-spots, where small regions have significantly higher numbers of crime than average. In addition, crimes may be selected by their Uniform Crime Reporting Code (see see Section~\ref{sec:data}), or into larger categories of crime including drugs and alcohol, sex crime, theft, or violent crime. Furthermore, a specific time period may be selected by specifying a date and time span (all data, a year, or a month).

The code violation data is sparser than the crime data (see Section~\ref{sec:data}), and a heat map would not be an appropriate visualization. Instead, these data are shown as circular pins on the map, with clusters of points represented by a larger pin with the number of points in the cluster inscribed. These data may also be selected by time in the same manner as the crime data. Community assets are visualized with pins as well, but without clustering as these data are sparse enough to not require aggregation. In addition, further information specific to the asset type is displayed on mouseover. 

The goal of this map is not to highlight areas of high crime in a negative manner, but to be a tool the community may use to coordinate their crime prevention strategies. For instance, an NPU public safety chair may be leading a drug prevention campaign. They may use the heat map to locate areas of drug use in their neighborhoods, and then overlay drug prevention programs from other sources on the map. They may then choose to concentrate their efforts on an area with drug usage and without another active drug prevention program. In another case, a public safety chair may use the map to examine historic data. They may zoom to a location where they have been active in the past and learn how crime has changed over time in the specific area that they work.

The aggregated data and correlation modules are generated with the Tableau data visualization software. The map is created with CartoDB, a database and geospatial visualization portal. The data are hosted with CartoDB's PostgreSQL server, and the visualization and navigational tools are created with the JQuery and CartoDB JavaScript libraries. All three modules will be hosted on the WCA data-dashboard with a release date in early August.

\section{Discussion}
\label{sec:discussion}
\textbf{Incorporating citizen data}\\ 
Our conversations with community residents have revealed an interest in a tool that would support citizen reports of crime. The underlying interest in such a tool appears to stem from community mistrust in the completeness and accuracy of public crime data. Including citizen data has the potential to empower local residents in ways that may be similar to citizen science efforts and the now-common social media documentation of crimes. However, including citizen data is fraught with its own issues of accuracy. At present, the Data Dashboard has no notion of users or access control. If citizen data were included, would that model need to change? What benefits and detriments would accrue if barriers to participation were included on portions of the web site? If citizens report crimes not included in public data sets, there is no clear way to substantiate the accuracy of the reports. How can citizen data be reflected in ways that respect the information yet do not endorse the accuracy? Does citizen crime reporting work with or against local law enforcement efforts? How can tools encourage positive community-police interactions, when and if that is appropriate? 

\textbf{Data literacy}\\
The success of tools like the Data Dashboard, and especially the Public Safety Portal, depend on data literacy. In nearly every discussion we had with stakeholders, the confusion between correlation and causation was abundantly clear. (Indeed, even we and other members of our Georgia Tech community easily confuse the two when looking at graphical data.) We see a need for teaching methods and tools that help explain and reinforce the difference between causation and correlation. Without this, community members risk taking away inaccurate information from data sets and limit their own ability to be successful advocates. Given how much confusion exists even among people with frequent exposure to data, we posit this is a key challenge for community and citizen based data efforts that will likely require creativity, local knowledge, data ambassadors, etc. 

\textbf{Asset-based approaches to data} \\
Community engagement experts often take an asset-based approach to development, where community knowledge, resources, aspirations and skills are emphasized over community deficits such as crime and poverty~\cite{mcknight2013basic}. The asset-based approach is challenging in settings such as the Data Dashboard project. The ``Public Safety'' Portal is so named as a step in an asset-based direction, rather than, for example, calling it the ``Crime'' Portal. Yet the data sets of key relevance to public safety are largely negative -- code violations, crime reports, demographic data such as education levels or income. We made several additional attempts to take an asset-based approach. We invited the community to suggest place-based assets that we could include on the map, in addition to schools and places of worship that we could easily identify. When we talk to community members, we include use cases that result in positive developments for the community, not just reduction in negative events. For example we show how data might be used to understand which community-based youth programs correlate with changes in crime. We do not yet know how effective our methods have been in conveying an asset-based approach, nor how those efforts interact with advocacy. 

\section{Conclusion and Future Work}
\label{sec:conclusion}
Once the first iteration of the Public Safety portal is complete, the WCA will lead a focus group with members of various groups including NPU public safety chairs, Atlanta police, juvenile justice activist, religious leaders, code enforcement and Department of Justice liaisons. This will be an opportunity for the WCA to hear how well the site meets the needs of the various organizations and what aspects need to be improved upon.  The WCA will use the comments to update the site before a full public launch in Fall 2016.  After the launch, the WCA will organize and lead several trainings to help guide community members on using the site.  These trainings will be open to the public and will be designed for a broad range of technical literacy.   

As the assets currently listed on the portal come from an agency perspective, the WCA plans in the future to work with community leaders to map neighborhood described assets.  These may be physical locations like basketball courts but also events like after school programs. Mapping crime and assets together will give organizations an opportunity to track the impact of their programs and make the case for expanding programs that have documented success. Expanding the list will also give a more robust description of the community and allow us to better understand how well we are incorporating an asset-based mindset, while also making accurate and important (negative) data available to the community.

\section{Acknowledgments}
We gratefully acknowledge the financial support provided to the Data Science for Social Good 2016 program from Oracle Academy and Georgia Tech. This work would not be possible without the participation of citizens in the Westside communities. We also thank Dr. Carl DiSalvo for his valuable comments. 

\nocite{*}
\bibliographystyle{abbrv}
\bibliography{references}

\end{document}